\documentclass[sigconf,preprint,nonacm]{acmart}
\makeatletter                   
\def\mdseries@tt{m}             
\makeatother                    

\usepackage[frozencache,cachedir=.]{minted}
\title{Generating Examples From CLI Usage: Can Transformers Help?}
 
\author{Roshanak Zilouchian}
\authornote{Equal Contribution}
\email{rozilouc@microsoft.com}
\affiliation{%
  \institution{Microsoft}
  \city{Redmond}
  \state{Washington}
  \country{USA}
}

\author{Spandan Garg$^*$}
\email{spgarg@microsoft.com}
\affiliation{%
 \institution{Microsoft}
 \city{Redmond}
 \state{Washington}
 \country{USA}}
 
\author{Colin B. Clement$^*$}
\email{coclemen@microsoft.com}
\affiliation{%
 \institution{Microsoft}
 \city{Redmond}
 \state{Washington}
 \country{USA}}

\author{Yevhen Mohylevskyy}
\email{yemohyle@microsoft.com}
\affiliation{%
 \institution{Microsoft}
 \city{Redmond}
 \state{Washington}
 \country{USA}}

\author{Neel Sundaresan}
\email{neels@microsoft.com}
\affiliation{%
 \institution{Microsoft}
 \city{Redmond}
 \state{Washington}
 \country{USA}}
\renewcommand{\shortauthors}{}

\begin{document}
\begin{abstract}
Continuous evolution in modern software often causes documentation, tutorials, and examples to be out of sync with changing interfaces and frameworks. Relying on outdated documentation and examples can lead programs to fail or be less efficient or even less secure. In response, programmers need to regularly turn to other resources on the web such as StackOverflow for examples to guide them in writing software. We recognize that this inconvenient, error-prone, and expensive process can be improved by using machine learning applied to software usage data. In this paper, we present our practical system which uses machine learning on large-scale telemetry data and documentation corpora, generating appropriate and complex examples that can be used to improve documentation. We discuss both feature-based and transformer-based machine learning approaches and demonstrate that our system achieves 100\% coverage for the used functionalities in the product, providing up-to-date examples upon every release and reduces the numbers of PRs submitted by software owners writing and editing documentation by $>$68\%. We also share valuable lessons learnt during the 3 years that our production quality system has been deployed for Azure Cloud Command Line Interface (Azure CLI).
\end{abstract}

\ccsdesc[500]{Computer systems organization~Embedded systems}
\ccsdesc[300]{Computer systems organization~Redundancy}
\ccsdesc{Computer systems organization~Robotics}
\ccsdesc[100]{Networks~Network reliability}

\keywords{Example Generation, Transformers, Software Documentation}


\maketitle
\thispagestyle{empty}
\pagestyle{plain}

\section{Introduction}
Modern software development involves continuous integration, deployment, and rapid releases. New frameworks, libraries, and APIs are created and the existing ones keep improving and changing. This rapid and constant change often presents a steep learning curve to developers. In many cases, spending time and effort to become proficient in using a library or API is not even productive as it may only be used a few times. Instead the most efficient way to guide developers is by providing code examples demonstrating how to use new APIs or interact with new frameworks~\cite{ko-vlhcc, forward-doceng}. An extensive survey of software developers has identified \emph{use of up-to-date examples} as one of the most important factors in useful documentation~\cite{forward-doceng}. However, documentation and code examples are usually added only as an afterthought to comply with regulations, often rendering them out of sync or incomplete~\cite{parnas,robillard}. Even when they exist, the documentation content and code examples are not updated in a timely manner~\cite{lethbridge}. Therefore, insufficient quantity and variation~\cite{robillard} in examples and incorrect examples~\cite{aghajani2020doc, Aghajani2019SoftwareDI} remain to be the major obstacles for developers learning to use an API.

Code examples shared in Blogs, Wikis and Q\&A sites have emerged as an alternative to supporting official documentation~\cite{pagano-msr, mamykina-chi}. However, such advice can go out of date in a matter of weeks. Further, when mining an enormous number of blogs and online articles, finding the most current or relevant examples can be difficult~\cite{robillard}. Additionally, blog articles or discussions on Q\&A sites are not officially maintained by the software owners and the examples may be of poor quality \cite{nasehi2012WhatMakesExample}. 

Knowledge discovery tools can address these challenges to some extent. Knowledge discovery tools provide recommendations in the form of code samples or artifacts~\cite{nykaza2002programmers, mclellan1998building}. However, they cannot offer help for uncommon code frameworks or when samples are not present, limiting their use as alternatives for missing documentation. To tackle these challenges, another line of research has emerged to augment documentation with synthesized examples~\cite{kim2009adding,montandon2013documenting, mar2011recommending}. Our work extends this line of prior work by generating up-to-date examples from usage data and other external sources of information and automatically inserting them into the official documentation.

Our example generation framework automatically creates and updates examples in software documentation upon every release. The examples generated by our platform have following qualities:
\begin{itemize}
    \item \textbf{Up-to-date examples} Our platform utilizes usage telemetry to generate new examples at every release cycle of a product, ensuring the examples are always up-to-date.
    
    \item \textbf{Representative of actual usage} Unlike bare-bones examples usually found in documentations that only cover basic scenarios, our examples are based on usage telemetry and, therefore, represent how current users use the software in practice.
    
     \item \textbf{Covering all used functionalities} Our automatically generated examples cover all used functionalities of the software, in contrast to human written examples which are usually provided for a few important functionalities.
\end{itemize}

Our example generation framework consists of two steps: (i) Identifying successful scenarios to build example templates based on prior user successes, and (ii) Translating the templates to human readable examples. For the second step, we experimented with a feature-based parameter type prediction model and a transformer-based neural parameter value generation model. We discuss the benefits and challenges of each model in a production environment.

Our example generation system has been deployed for Azure Command Line Interface (Azure CLI), a large scale, open-source cloud developer command line environment. Our comparative study between our generated examples and the human written examples by software owners showed that our examples can help developers by covering all active features with a higher quality than the software owner’s examples. In addition, we found that our example generation pipeline was able to reduce the number of PRs submitted by software owners to write or edit documentation by $>$68\%.

In this paper we make the following contributions:
\begin{enumerate}
    \item we present a production-quality example generation platform which can generate up-to-date examples that cover all used functionalities,
    \item discuss the benefits and challenges of a neural model and a feature-based model in a production environment,
    \item share lessons learned from the deployment of our example generation platform in production.
\end{enumerate}

\section{Related Work}
Prior work has tackled the problems posed by rapidly changing APIs and frameworks in software development~\cite{robins2003learning} in different ways: crowd-sourced documentation, augmenting documentation with examples, and knowledge discovery tools.
\subsection{Crowd-Sourced Documentation}
As the leading way to learn about new features and APIs, web search enables developers to discover socially-mediated sources of information in addition to official documentation. Blogs, wikis and Q\&A sites are commonly used to complement the official documentation. A study of Google search results on jQuery API showed that at least one blog post and StackOverflow question appear on the first page of the search results for 84\% of methods in jQuery \cite{parnin2011measuring}. 

However, it is not clear whether some of these additional sources will resolve staleness or the lack of examples in official documentations. For example, a study on blogging behaviors of developers has revealed that only 1.8\% of relevant blog posts contain source code \cite{pagano-msr}. This means that developers use blogs mainly to communicate and coordinate functional requirements as opposed to documenting code. Similarly, studies of Q\&A websites such as StackOverflow have shown some software tools or APIs may not get enough coverage on StackOverflow \cite{parnin2012crowd}. Even for popular software tools, the coverage accumulates very slowly. For instance, for Android API classes the coverage after one year was only 30\% \cite{parnin2012crowd}. This coverage is much worse in specialized software tools. Also, even questions posted to StackOverflow for popular software systems are usually answered by a small group of experts; such experts are hard to find for systems with smaller communities. Failure to find experts has been identified as one of the key reasons for unanswered questions on StackOverflow~\cite{asaduzzaman2013answering}. Our work fills the coverage and staleness gap in documentation by generating up-to-date examples based on usage for all of used commands and APIs.

\subsection{Augmenting Documentation with Examples}
Prior research has identified examples as a key learning resource in software development~\cite{nykaza2002programmers, mclellan1998building, holmes2009end}. \citet{kim2009adding} proposes a technique to extract code examples and integrate the examples into API documentation. \citet{montandon2013documenting} describes APIMiner, a platform which extracts code examples from software repositories and instruments the standard Java API documentation with code examples. PorpER-Doc is another tool which accepts queries from API developers and suggests proper code examples for documentation purposes~\cite{mar2011recommending}. 
\citet{Buse2012SynthesizingAU} presents a technique for automatically synthesizing human-readable API usage examples. Our work extends these works by generating examples from usage data and mining public resources, automatically inserting the examples into official documentation.

\subsection{Knowledge Discovery Tools}
Knowledge discovery tools can come to the rescue when there are stale examples in API and framework documentation. For instance, eMoose highlights rules or caveats of API calls in the documentation~\cite{dekel2009improving}. 
XSnippet uses the code context such as types of methods and variables to locate sample code for object instantiation~\cite{sahavechaphan2006xsnippet}. Similarly,  PARSEWeb~\cite{ThummalapentaICACE2007} and Prospector~\cite{mandelin2005jungloid} are also designed to provide examples of object instantiation to help developers navigate complex APIs. 
While clearly filling a niche, these tools have been found to be limited in their scope: they cannot offer help when code samples are not present or certain API calls have not been widely used. Our work ameliorates this limitation by creating high quality examples demonstrating how to use a tool or framework from previously successful usages.    

\section{Azure CLI}
While our example generation platform can be leveraged for any application where usage data is available, for the purpose of this paper, we will specifically target a popular Command Line Interface (CLI) that is used to interact with the Microsoft Azure cloud platform, referred to as Azure CLI in this paper. Figure~\ref{fig:cli_example} shows an example of an Azure CLI command which creates a virtual machine.
\begin{figure}[h]
    \begin{minted}[fontsize=\footnotesize]{bash}
    az vm create --image UbuntuLTS --admin-username azureuser
      --name MyVM --ssh-key-value ~/.ssh/id_rsa.pub 
      --resource-group MyResourceGroup --location westeurope
    \end{minted}
    \caption{An example of an Azure CLI command for creating an Ubuntu virtual machine on Azure. All Azure CLI commands begin with `az' followed by a command name. A command may be followed by a set of parameters and corresponding values, where parameters begin with `-{}-'.}
    \label{fig:cli_example}
\end{figure} 
Each Azure CLI command consists of a command name (e.g. \mintinline{bash}{az vm create}) and a set of parameters names which usually start with a \mintinline{bash}{--} flag (e.g. \mintinline{bash}{--image}) and are followed by a parameter value (e.g. \mintinline{bash}{UbuntuLTS}). Overall, Azure CLI has more than 3600 commands. On a monthly basis, users run millions of commands to create, delete, and manage resources on Azure. While many of these commands run successfully, failures are quite common. A command may fail for various reasons like incorrect parameter combinations, errors in parameter names or parameter values, wrong assumptions about the state of a resource, or even service problems. In Azure CLI, user faults which includes wrong parameters combinations and errors in parameter names or values can account for up to 22\% of command failures. These errors occur mainly due to lack of documentation and examples covering various parameter combinations. For instance, each Azure CLI command has at most 76 parameters and on average 10 parameters. The average number of parameters specified by users is 4, while the average number of parameters in the examples provided in the official documentation is 1. Therefore, the examples provided in the documentation likely do not fully capture the way Azure CLI is used in practice. This potential gap between official documentation and the actual usage of Azure CLI will only grow larger in time, and can cost both companies and customers a significant amount in wasted time and resources. 

\section{Example Template Generation}
\begin{figure}[h]
    \includegraphics[width=8cm]{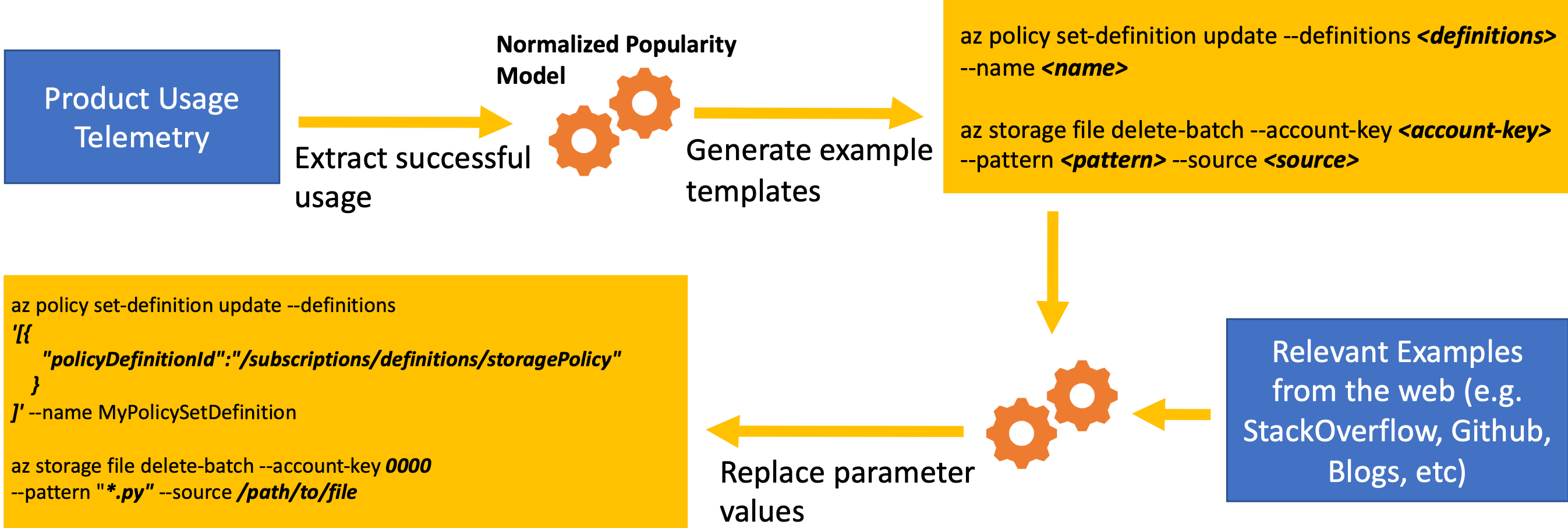}
    \caption{Overview of our example generation framework. We use product usage telemetry to generate example templates. We then collect relevant examples from various web sources and use them to train models that can find or generate the best parameter values for each parameter. Finally, the parameter values are added to the template giving us the resulting examples.}
   \label{fig:arch}
\end{figure}

Our example generation framework consists of two steps: (i) identifying successful scenarios to build example templates based on prior user successes, and (ii) translating templates into human readable examples. Figure \ref{fig:arch} shows an overview of our pipeline.

In order to identify successful scenarios, we analyze the usage telemetry of Azure CLI. This telemetry data includes the CLI commands, a list of parameters used with each command, and whether the execution of the command was successful or not. Keeping customer privacy in mind, the usage telemetry data \emph{does not} include the concrete parameter values, preventing potentially private information like user-name or email addresses from leaking into the machine learning model and possibly into the examples. 

For each upcoming release of Azure CLI, we collect around \emph{3.20 billion} successful commands which were executed for the last three months prior to the release. We then remove the commands corresponding to the old version and all the help calls, which did not result in an actual command execution from the data. This leaves us with $\sim$\emph{3.19 billion} successful command and parameter set pairs. We then sort the unique command and parameter set pairs based on frequency of unique users. Going through the list of all parameter sets for all commands, we then take the top three most frequent parameter sets for each command to build up to three example templates. Since we do not have the values of parameters in the usage telemetry, we use a placeholder value based on the parameter name in the generated templates (e.g. \mintinline{bash}{<image>} for a parameter named \mintinline{bash}{--image}). Figure \ref{fig:template_example} shows an example of a template generated for the virtual machine (VM) creation command with placeholders. 

\begin{figure}[h]
    \begin{minted}[fontsize=\footnotesize]{bash}
    az vm create --image <image> --admin-username <admin-username>
      --name <name> --ssh-key-value <ssh-key-value> 
      --resource-group <resource-group> --location <location>
    \end{minted}
    \caption{An example of a template created from usage record of \mintinline{bash}{az vm create} command. The parameter values are replaced with placeholders, which are parameter names surrounded by angle brackets, e.g. \mintinline{bash}{<ParameterName>}.
    }
    \label{fig:template_example}
\end{figure}

\section{Parameter value Generation}
An example is more useful if its parameter values are concrete and not placeholders as they give users more information about acceptable values, value formats (e.g. for date/time), and share common conventions. Here is an example of an Azure CLI command which shows how to update an Azure application with placeholders: 
\begin{minted}[fontsize=\scriptsize,escapeinside=||]{py}
    az ad app update --id |\bfseries{<id>}| --start-date |\bfseries{<start-date>}|
\end{minted}
Contrast this with an example containing actual values:
\begin{minted}[fontsize=\scriptsize,escapeinside=//]{py}
    az ad app update --id /\bfseries{e042ec-34cd-498f-9d9f-14567814}/
                     --start-date /\bfseries{/"2017-01-01"/}/
\end{minted}
where \mintinline{bash}{--id} is, thus, understood to take an alphanumeric GUID and \mintinline{bash}{--start-date} an ISO-formatted date string. 

In order to replace the placeholders with actual values, we developed two models: (i) a feature-based parameter type prediction model, and (ii) a neural parameter value generation model.

Our feature-based parameter type prediction model predicts the parameter's type first. It then uses the identified type to choose a correct value from a pre-computed lookup table of collected values for a given parameter. On the other hand, our neural parameter value generation model receives an example template as an input and generates parameter values. We now explain the data we used and the model training details.

\subsection{Data Collection}
\label{data-collection}
While the usage telemetry data was enough to create example templates, it lacked parameter values. Therefore, we needed Azure CLI examples with parameter values to train our parameter value generation models. To find these examples, we first collected the following documents:
\begin{itemize}
    \item All questions and answer posts from StackOverflow, which were at most one year old and were tagged with `Azure' or `Azure-CLI' for a total of 1481 posts.
    \item All 9167 GitHub issues submitted to Azure CLI's repository.
    \item All $\sim$14k pages of official Azure blogs and documentations. 
\end{itemize}
We then developed a parser to identify Azure CLI commands from the collected documents. The parser looks for code blocks starting with \mintinline{bash}{az <command>} or code blocks, which are tagged with an \emph{azure-cli} language tag, yielding $>$22K Azure CLI examples. We then filtered out the examples that would only run on Azure CLI versions released before January 2019. We also filtered out examples that had invalid commands or parameter names, values featuring typos, or values affected by breaking changes in new releases. After filtering, we were left with $\sim$7K unique and syntactically correct examples.

\subsection{Feature-based Parameter Type Prediction Model}
For our feature-based parameter type prediction model we hand-labeled the parameters in the final dataset of 7K examples into 15 categories based on the types of acceptable values. These categories were also verified with the software owners. Table \ref{tab:table1} shows a list of these categories. For each command and parameter in our dataset we also retrieved the command and parameter descriptions from Azure CLI's documentation. We then cast our data into features vectors and trained a classifier to predict the parameter types.

\begin{table}[htbp]
\centering 
\footnotesize
 \caption{Azure CLI parameter types and their respective frequencies in collected data.}
 \label{tab:table1}
\begin{tabular}{l l} 
 \hline
 Category & Frequency \\ [0.5ex] 
 \hline\hline
 String & 5228 \\ 
 \hline
 Enum & 713 \\
  \hline
 Integer & 273\\
 \hline
 GUID & 246 \\
 \hline
Folder/File Path & 241\\
 \hline
 Command Specific/Unknown & 201 \\
 \hline
 IP-Address & 196 \\
 \hline
URL/E-Mail & 166\\
 \hline
Build Info	& 131 \\ 
 \hline
Quoted Strings & 125\\
 \hline
Version & 45\\
 \hline
Time/Duration & 23\\
 \hline
Keys/Tokens	 & 14\\
 \hline
Int With Specific Format	 & 6\\
 \hline
Permission Formats & 5\\
 \hline
\end{tabular}
\end{table}

\subsubsection{Feature Embeddings}
Our raw features include the \emph{command} name, the \emph{parameter} name, the name of the \emph{module} containing the command, the \emph{parameter description} in the Azure documentation, and the \emph{command description} from the Azure documentation. 
We performed several pre-processing steps on the text of each feature. We first transformed the text to lower-case, removed all the special non-ASCII characters and common stop words. We then performed WordNet-based lemmatization over the words, which is the removal of inflectional endings of words, replacing them with their base, known as the lemma, reducing the necessary vocabulary of the feature vectors. We convert each sequence of words in our features to a vector representation using a bag-of-words representation~\cite{manning1999foundations}. For \emph{parameter name}, \emph{command name}, and \emph{module name} the traditional bag-of-words worked well because these features have a small vocabulary (<100 words) and, therefore, we did not have to limit the size of our feature vector. The other two features, \emph{parameter description} and \emph{command description}, include several sentences with a lot of variation in word usage and, as a result, a large vocabulary. To limit the vocabulary size, we selected the top 75 words for each parameter type category based on its correlation with the category. 
\begin{figure}[h]
    \includegraphics[width=5.5cm]{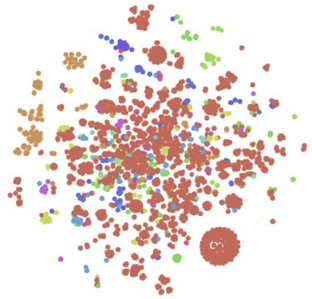}
    \caption{We used t-SNE to visually map each high-dimensional point in our data set to two dimensional space. Each point is color coded based on the type of the parameter it represents. String points (Red) tend to overlap with minority class examples.}
    \label{fig:tsne1}
\end{figure}
\begin{figure}[h]
    \includegraphics[width=6.5cm]{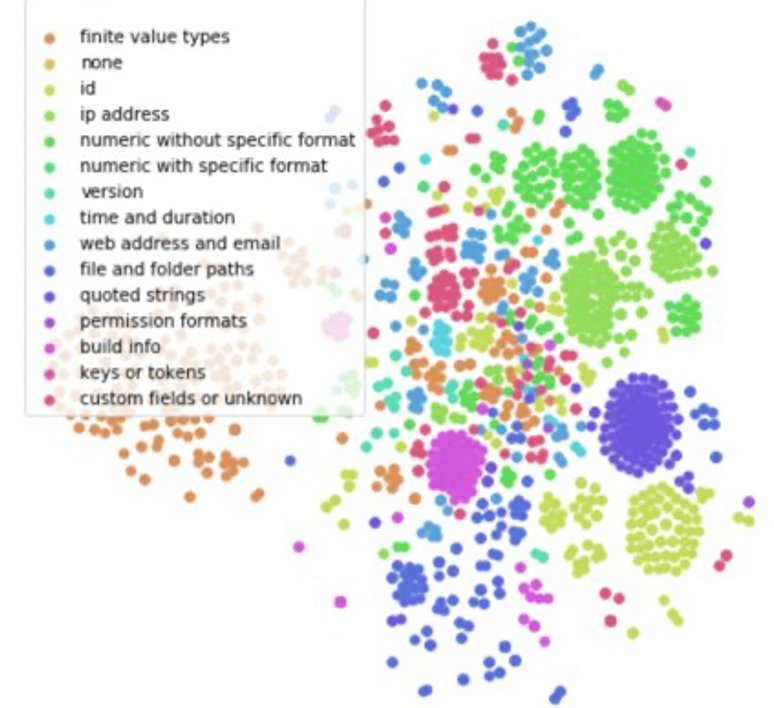}
    \caption{We used t-SNE to visually map each high-dimensional point in our data set to two dimensional space. Each point is color coded based on the type of the parameter it represents. After removing all `string' parameter types from the data, the t-SNE graph shows a clear separation of all other minority classes in the two dimensional space. }
    \label{fig:tsne2}
\end{figure}
\subsubsection{Classifier}
Using the features, we trained a Random Forest classifier to predict the type of the parameters. Our data set had a data imbalance issue as the majority of the parameters were of the type `string'. We visualized our data using t-SNE \cite{maaten2008visualizing}, which maps each high-dimensional data point to a location in a two-dimensional map. In the t-SNE graph of our data set we observed that the points in the graph representing the `string' class overlap with points from other minority classes at every value of perplexity we tried (Figure \ref{fig:tsne1}).
Removing `string' points entirely led to a clear separation of minority classes in the t-SNE graph (Figure \ref{fig:tsne2}).
Therefore, we decided to use two classifiers: (i) a `string' vs `non-string' classifier and (ii) a type classifier for classifying `non-string' examples into their finer types. For both classifiers, Random Forest yielded the best results when we experimented with various classification algorithms.
 
\subsubsection{Results}
Table \ref{tab:table3} and \ref{tab:table4} show the precision-recall values we achieved for the `string' classifier using bag-of-words features and the `non-string' finer type classifiers respectively. As shown in the tables our classifier has high F-1 score for the majority of classes.  
\begin{table}[h!]
\centering 
\footnotesize
 \caption{Precision/recall of the string classifier with a 3-fold cross validation.}
 \label{tab:table3}
\begin{tabular}{l l l l l} 
 \hline
 & Precision & Recall & F-1 Score & Support \\ [0.5ex] 
 \hline\hline
 String	& 1.00 &	0.86 &	0.92 &	5228 \\
 \hline
Non-String &	0.76 &	1.00 &	0.86 &	2385\\
 \hline\hline
Weighted Avg. &	0.92 &	0.90 &	0.90 &	7613\\
\hline
\end{tabular}
\end{table}

\begin{table}[h!]
\centering 
\footnotesize
 \caption{Precision/recall of a bag-of-words based parameter type classifier with 3-fold cross validation.}
 \label{tab:table4}
\begin{tabular}{l l l l l} 
 \hline
 Category & Precision & Recall & F-1 Score & Support \\ [0.5ex] 
 \hline\hline
Enum	& 0.89 &	0.98 &	0.94 &	713\\
\hline
Integer &	0.89 &	0.88 &	0.88 &	273\\
\hline
GUID &	0.89 &	0.77 &	0.82 &	246\\
\hline
Folder/File Path &	0.94 &	0.95 &	0.95 &	241\\
\hline
Command Specific &	0.79 &	0.72 &	0.75 &	201\\
\hline
IP-Address &	1.00 &	0.84 &	0.91 &	196\\
\hline
URL/E-Mail &	0.98 &	1.00 &	0.99 &	166\\
\hline
Build Info &	0.99 &	1.00 &	1.00 &	131\\
\hline
Quoted Strings &	0.67 &	0.90 &	0.76 &	125\\
\hline
Version	& 0.87 &	0.29 &	0.43 &	45\\
\hline
Time/Duration &	1.00 &	0.70 &	0.82 &	23\\
\hline
Int With Format &	1.00 &	1.00 &	1.00 &	6\\
\hline
Permissions &	1.00 &	1.00 &	1.00 &	5\\
\hline
Keys/Tokens &	0.93 &	1.00 &	0.97 &	14\\
\hline
\hline				
Weighted Avg. &	0.90 &	0.89 & 0.89	& 2385\\
\hline
\end{tabular}
\end{table}

\subsubsection{Parameter Value Lookup}
We use the values from our collected examples (explained in sec.~\ref{data-collection}) to build a lookup table of possible values for each parameter. We then use regular expressions to make sure the collected values in the lookup table have proper syntax for the parameter's predicted type (IP-Address, File Path, etc.). For the `string' category, we use the parameter description in the documentation to create a valid name. For example, if the description of the parameter was "Name of the web app.", we use a regex to generate MyWebApp as the value for the name. If the lookup table doesn't include a type-correct value for a parameter, we retain the placeholder value from the template. 

\subsection{Neural Parameter Value Generation Model}
Transformers are a family of neural networks which currently obtain state of the art results for applications in natural language processing (NLP) such as machine translation, question answering, or document summarization~\cite{vaswani2017attention}. Since the introduction of transformers in 2017, several variations of transformer models have been developed including BERT \cite{devlin2018bert}, RoBERTa \cite{Liu2019RoBERTaAR}, and BART \cite{lewis2020bart} among others. These models are usually trained on a large amount of unlabeled data and then fine-tuned on a smaller task specific set of labeled data for a particular downstream task. 

We decided to experiment with a neural model because of several practical advantages of such models including (i) lower maintenance cost as these models need to be fine-tuned on more data over-time as opposed to feature-based models that usually need major feature engineering updates. (ii) a neural model pipeline enables us to experiment with other down-stream tasks to provide ML based solutions for other future scenarios such as command completion. (iii) the majority of research and development in NLP is focused on neural models, therefore using a neural model enables us to easily adopt the state of the art models for our down stream tasks.

In this work, we leverage from BART's architecture which combines Bidirectional and Auto-Regressive Transformers \cite{lewis2020bart}. For pretraining, the input text is corrupted with an arbitrary noising function and the model is trained to reconstruct the corrupted text.

\subsubsection{Pretraining}
Prior work in leveraging transformers for code completion has shown that pretraining on code snippets can significantly improve model performance on specific tasks such as method and docstring prediction\cite{clement2020pymt5}. Inspired by the prior work, we pretrained sequence-to-sequence transformers using a span-masking objecting~\cite{lewis2020bart} on publicly available shell script data. The span-masking objective essentially replaces random spans of input tokens with a \texttt{<MASK>} token, and the model is trained to predict all the tokens replaced by the mask, separated by mask tokens.

For pretraining, we collected 51K GitHub repositories with $\geq$5 stars that were composed primarily of shell scripts, resulting in 328K unique scripts with 54 million total lines of code. We then pretrained our 139M and 406M parameter transformers (BART-base and BART-large, respectively) on this corpus for 60 epochs on four Nvidia Tesla V100 16GB GPUs, $\sim$48 GPU-hours total for the larger model.

\subsubsection{Fine-Tuning}
For fine-tuning we used the 7k unique examples collected from the web (explained in \ref{data-collection}). We fine-tuned our shell-pretrained transformer models for predicting Azure CLI parameter values by replacing each sub-sequence of parameter value tokens with a \texttt{<MASK>} token, and training the model to predict tokens for each parameter value, separated by mask tokens. In this way, the model is taught to allocate any number of tokens to each parameter value. We call the resulting parameter-prediction models DeepDevAZ and DeepDevAZ-large.

\subsubsection{Data Augmentation}
Our fine-tuning data was not large by modern deep learning standards, as we only had about 7000 unique Azure CLI commands. In order to improve the model training we augmented the data by adding copies of each command with all permutations of masking and unmasking. For example, a given command with two parameters yielded 3 examples for training, as we masked both, and one parameter, and then the second parameter. In general this yields $2^n-1$ copies for a command with $n$ parameters. This also improves the range of tasks DeepDevAZ can complete, allowing complete or partial parameter naming.

\section{Experiments}
We perform two experiments to gauge the effectiveness of our models. The first experiment focuses on comparing the neural parameter generation model with other baselines and the second experiment compares the feature-based and the neural generation approach for replacing placeholder values in our example templates.

\subsection{Experiment 1: Comparing neural approaches}
We compared our DeepDevAZ and DeepDevAZ-large models with two baseline models:  (i) a RoBERTa model pre-trained on english and fine tuned on our examples data set (RoBERTa-ENG-AZ) with token masking objective and (ii) a BART model pre-trained on english and fine-tuned on our examples data set (BART-ENG-AZ) with span masking objective. We use ROUGE-1, ROUGE-2 and ROUGE-L \cite{RougeACL2004} metrics for this evaluation. Table \ref{table:exp1} shows the scores achieved by our DeepDevAZ model compared to the baselines.

The substantial difference between our RoBERTa-ENG-AZ baseline, which uses a BERT architecture and the other models that use BART, indicates the advantage of task-specific training objectives. RoBERTa-ENG-AZ is trained on the masked language modeling task, and decoding parameter values, which are composed of multiple tokens, requires an iterative process of infilling mask tokens, which is not how the model was trained. The sequence-to-sequence models enable an in-filling scheme, where arbitrary length spans of text are replaced with a single mask token while BERT can only predict one masked token. Therefore the BART-style sequence-to-sequence model is more appropriate for parameter value generation where parameter values usually consist of more than one token. 

Comparing sequence-to-sequence models pre-trained on english and shell script data, we observe that the publicly released (406M parameter) BART-large checkpoint pre-trained on English performs slightly better than our smaller (139M parameter) DeepDevAZ, but our (406M parameter) DeepDevAz-large model is the best model overall. Therefore, we conclude that large model size is advantageous even in this small data regime, and pre-training on Shell scripts is more valuable than pre-training on English alone.

\begin{table}[htbp]
    \centering
    \footnotesize
    \caption{Performance of DeepDevAZ model and the other two baselines.}
    \label{table:exp1}
    \begin{tabular}{l l l l l}
        Model & Stat. & R1 & R2 & RL  \\\hline\hline
        RoBERTa-ENG-AZ &  Prec. & 15 & 1.4 & 18 \\
        & Rec.& 10 & 1 & 12 \\
        & F1 & 12 & 1.1 & 14 \\\hline
        BART-large &  Prec. & 51.3 & 30.6 & 30.7 \\
        (english pretrained) & Rec.& 51.0 & 30.7 & 51.5 \\
        & F1 & 51.1 & 30.6 & 51.1 \\\hline
        DeepDevAZ &  Prec. & 44.2 & 26.6 & 44.0 \\
        & Rec.& 47.7 & 28.6 & 49.4 \\
        & F1 & 45.5 & 27.4 & 46.1 \\\hline
        DeepDevAZ-large &  Prec. & \bf 55.1 & \bf 35.1 & \bf 55.0 \\
        & Rec.& \bf 54.7 & \bf 35.0 & \bf 55.9 \\
        & F1 & \bf 54.8 & \bf 35.0 & \bf 55.2 \\\hline
    \end{tabular}
\end{table}

\subsection{Experiment 2: Comparing neural and feature-based models}
We leveraged ROUGE as a metric in our first experiment as it provides an efficient way to compare large numbers of predictions produced by various neural models. However, prior research has shown the shortcomings of ROUGE as a metric, which causes it to correlate poorly with human judgment \cite{liu2016not, novikova2017we}. To fill this gap, we performed a human judgement evaluation comparing the examples our DeepDevAZ-large model has produced with examples produced by our feature-based model for the 100 most frequently used Azure CLI commands. This evaluation was performed by two of the authors, who are knowledgeable in Azure CLI, with help from domain experts. The examples were evaluated for their syntactical correctness and how likely they were to be written by a human. For verifying syntactic correctness, an automated execution of the produced examples was insufficient for two main reasons. First, some of these examples rely on other resources to already exist in Azure in order to execute correctly. Second, some generated examples have placeholder values that may be syntactically correct, but will not execute without replacing placeholders with real values. Aside from syntactical correctness, we also verified human readability. For instance, predicting a value such as "mymymy" for a virtual machine name may be syntactically correct, but it is not a value an actual developer will pick. To this end, the authors collaborated with 3 domain experts to determine if examples satisfy human readability. Table \ref{tab:human-eval} shows the results of these comparisons.


\begin{table}[htbp]
    \centering
    \scriptsize
    \caption{Human evaluation of 100 frequent Azure CLI commands comparing the examples generated by our feature-based and neural models. The evaluation was performed both on syntactical correctness of the example as well as how likely the example is to be written by human.}
    \begin{tabular}{p{0.37\linewidth} p{0.15\linewidth} p{0.28\linewidth}}
        Model & Judged correct & Non-placeholder examples  \\\hline\hline
        Feature-based Parameter Prediction &  99 & 87 \\\hline
        DeepDevAZ-large &  87 & 97 \\\hline
    \end{tabular}
    \label{tab:human-eval}
\end{table}
The evaluation showed that majority of the examples generated by our feature-based model are syntactically correct. However, they also include a lot more placeholders in comparison to the neural model, which caused the examples with placeholders to not appear likely to be written by human. Our feature-based model uses placeholder values when type-correct values do not exist in the lookup table. Although the resulting examples are not judged as incorrect, they are not as useful as human-written examples, which usually contain concrete parameter values. Another challenge with our feature-based model is its inability to consider correlations between parameter values when choosing a value for a specific parameter. For instance, the following example generated by the feature-based model for \mintinline{bash}{az resource show} is incorrect:
\begin{minted}[fontsize=\scriptsize,escapeinside=||]{py}
  az resource show --name |\bfseries{MySubnet}| --resource-group MyResourceGroup
    --resource-type "Microsoft.Compute/virtualMachines"
\end{minted}
While the type of the resource is a virtual machine, the name that has been chosen is clearly a subnetwork name. Therefore this example is semantically incorrect and can confuse the users. 

In contrast our neural model generates a correct example: 
\begin{minted}[fontsize=\scriptsize,escapeinside=||]{py}
  az resource show --name |\bfseries{MyVM}| --resource-group MyResourceGroup
    --resource-type "Microsoft.Compute/virtualMachines"
\end{minted}
This is because unlike the feature-based model, our neural model considers the command and \emph{all of the} parameters into account when generating values for a parameter.


DeepDevAZ makes a few more mistakes than the feature-based model, majority of which are dominated by commands which have no example parameters in our training corpus. Whereas, the feature-based model chooses an anodyne placeholder for these missing examples, DeepDevAZ attempts to be creative, producing somewhat spurious, unconstrained results. 
The parameters where the DeepDevAZ model fails to generate a correct value for are usually complex in nature. For instance, in one example it fails to generate a correct value for a database partition key and in another it predicts the role assignment value for a data warehouse incorrectly.


Examining the correct examples our neural model generates, we observe that the neural model is learning and generating examples similar to what humans write. For instance, our neural model was able to generate the following example:
\begin{minted}[fontsize=\scriptsize]{bash}
    az storage share-rm delete --storage-account MyStorageAccount
        --name MyShare
\end{minted}
As we can see, the model is learning to correctly associate storage shares with names like "MyShare", similarly with storage account. 

Similar examples exist where our neural model is able to generate correct values for a variety of parameter types such as IP-address, file-path, data-time, etc. While the neural model fails to generate values for some of the complex parameters that it hasn't seen before, the fact that it correctly generates values for a wide range of parameters invites for future investments in the neural approach.


Below we explain how we deployed and experimented with these models in production and how our automated examples affected Azure CLI documentation in action.
\begin{figure}[h]
    \includegraphics[width=8cm]{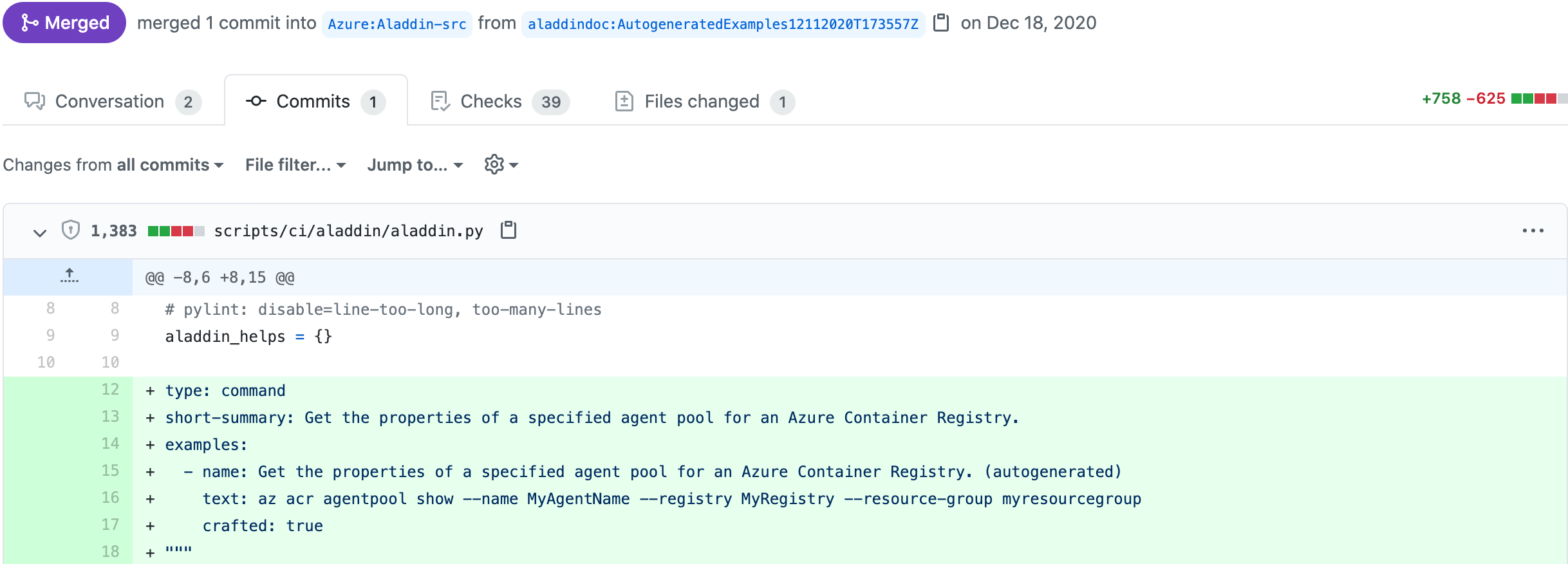}
    \caption{Commit from a Pull Request (PR) that was automatically generated and submitted to Azure CLI’s GitHub repo showing how examples being added to various services. Our example generation platform connects to an automatic PR generation module that creates PRs to add our generated the examples to Azure CLI on every release.}
    \label{fig:pr}
\end{figure}

\section{Deploying in Production}
\begin{figure}[htbp]
    \includegraphics[width=8cm]{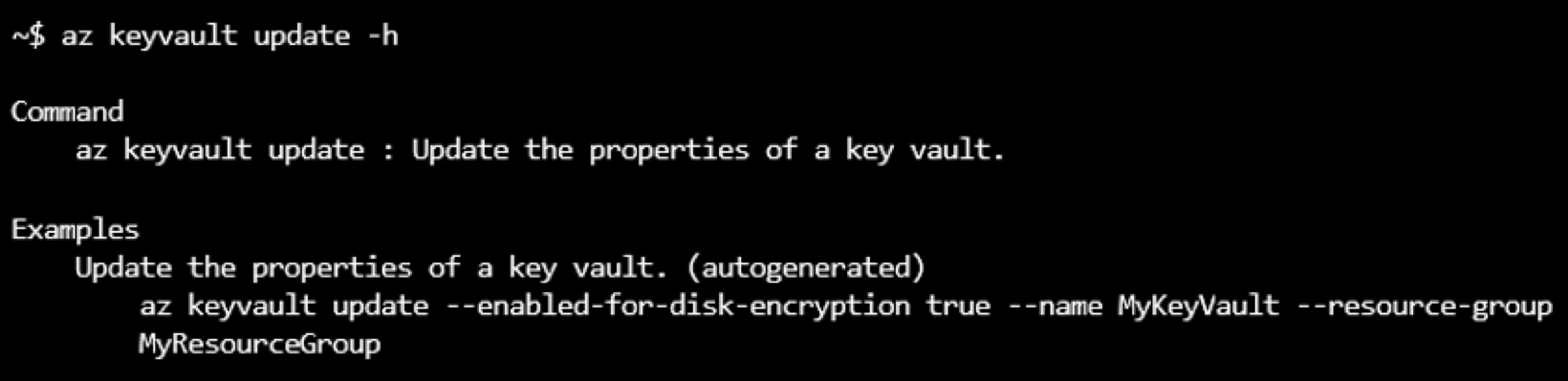}
    \caption{Examples and help content are accessible through the command line by using \mintinline{bash}{--help}/\mintinline{bash}{-h} with the command. In this figure, the user has called help on ‘az keyvault update’.}
    \label{fig:help}
\end{figure}

To evaluate the effectiveness of our example generation platform in a real practical setting, we connected our example generation platform to an automatic Pull Request (PR) generation module. This module submits Pull Requests to insert our examples into the official Azure CLI documentation on each product release cycle. A PR is a method of submitting code contributions to a code base. A developer submits a PR when they want to incorporate their code contributions/changes into a code base after a code review by one or more developers. Figure~\ref{fig:pr} shows an example of a PR that adds our example to the current Azure CLI documentation. Once integrated in the code base, developers can access the examples both through the command line help by typing the command name followed by \mintinline{bash}{--help} or \mintinline{bash}{-h} in the command line (fig.~\ref{fig:help}). Alternatively, they can view the examples on the online reference documentation (fig.\ref{fig:docs_compare}). To evaluate the effectiveness of our example generation platform in action, we examined the coverage and quality of the live examples.

\subsection{Coverage of Examples}
\begin{figure}[htbp]
    \includegraphics[width=8cm]{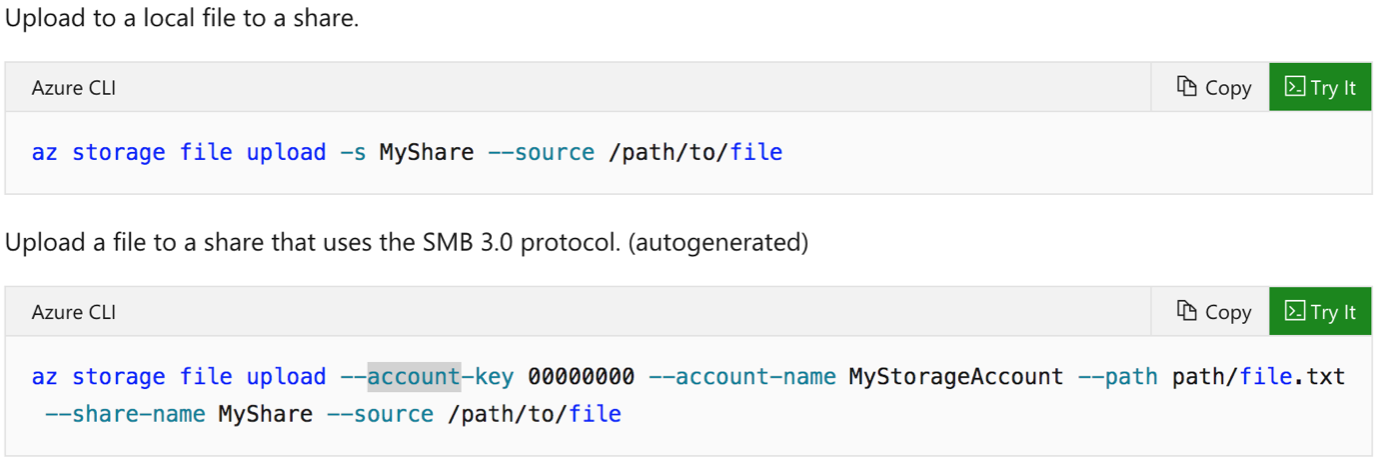}
    \caption{Screenshot of two examples from the Azure CLI documentation. The first one is added by the software owners and covers the basic case. While the second one is added by our platform (tagged with `autogenerated') and covers a more complex scenario showcasing a broader parameter set.}
    \label{fig:docs_compare}
\end{figure}

We first examined the coverage and quality of our generated examples. We observed that the examples written by software owners (human-written examples) cover only 55\% of the commands in Azure CLI. This means that software-owner-added examples account for a little over half of the Azure CLI commands, while our generated examples (machine generated examples) cover 100\% of the commands. This means that we can achieve algorithmically a scale of coverage that is difficult to achieve through manually written of examples. Additionally, while human-written examples on average cover only 20\% of the parameters for a given command, our machine-generated examples cover 32\%. Therefore, machine-generated examples not only cover more commands, they also cover more service functionalities and scenarios in Azure. In summary, we see an improvement of 82\% in command coverage and 60\% in parameter coverage compared to human-written examples.
Figure~\ref{fig:docs_compare} shows a screenshot of two examples for the same command in Azure CLI documentation. While the human written example on top covers a simple use-case of the command with only the required parameters, our machine generated one on the bottom (tagged with an `autogenerated' flag) supports a more complex scenario, involving more parameters. 

\subsection{Quality of Examples}
Besides coverage, we study how the quality of our machine-generated examples compares to human-written examples. As mentioned before, one of the primary ways of accessing examples in Azure CLI is through a help call on a command (invoking a command with \mintinline{bash}{--help} or \mintinline{bash}{-h}). These help calls are usually followed by an actual usage of the command with the user's desired set of parameters. This usage call following help should be successful, if the documentation and examples displayed in the help were useful to the user. Therefore, we can associate help calls with consecutive command usage calls immediately following it, within the same usage session. We take the success rate of the usage calls following the help calls as an approximate measure of quality. Since our machine-generated examples were added to a certain version of Azure CLI (version 60), we have a clean experiment comparing help success before and after the introduction of our generated examples.

\begin{figure}[htbp]
    \includegraphics[width=7.5cm]{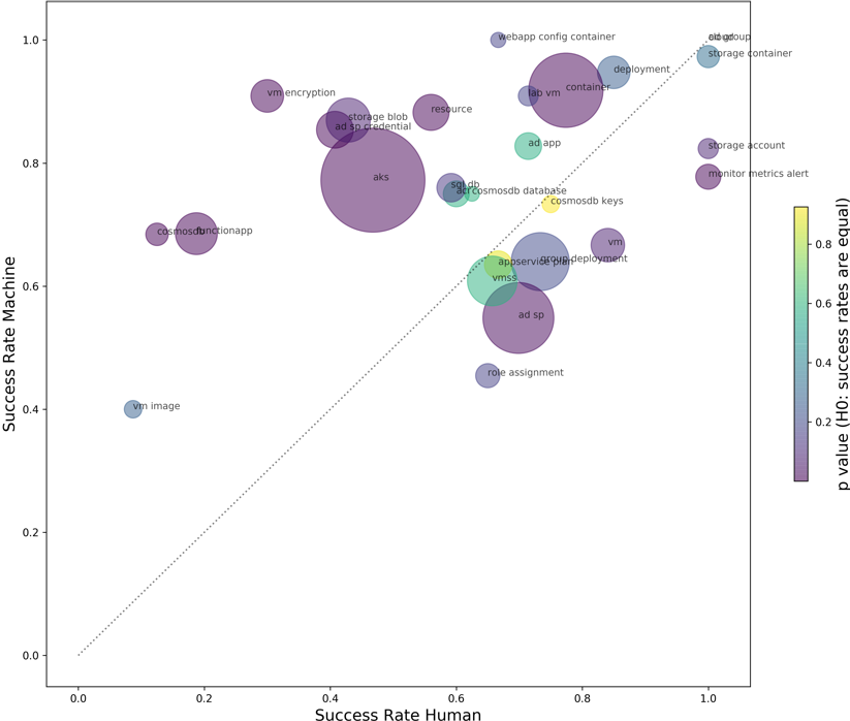}
    \caption{Success rates of machine-generated vs human-written examples after a help call.}
    \label{fig:success_rate}
\end{figure}

Figure~\ref{fig:success_rate} shows a plot of the before-mentioned quality metric. We first group commands into "command groups", which are groups of commands targeting similar resources in Azure. Each command group is represented by a bubble on the plot. For each command group, we compute the success rates of usages following the help call, where the command usage matches the parameter set shown in a human-written or machine-generated example. These rates correspond to abscissa and ordinate, respectively. The bubble size represents the customer usage of such commands over a period of 30 days (including both types of examples).


If the bubbles are spread along the diagonal, the human written examples and the machine generated examples, found in the output of \mintinline{bash}{--help}, are equally successful, while skewing of bubble density towards either the lower right or upper left corner would suggest the corresponding examples to be more effective.  We compute $p$-value of the observed deviations off the diagonal under the null hypothesis that the examples are equally effective. The $p$-values are encoded by the color on the plot, where darker colors are more significant than lighter ones. We can observe that, for the majority of command groups, our machine-generated examples are more helpful than the human-written ones.

\subsection{Software Owners’ Workflow}
Finally, we analyze the impact of our example generation pipeline on software owners’ workload. Our analysis reveals that our example generation platform saves Azure CLI developers a significant amount of time spent writing/editing reference docs. For example, in 2018 (before deployment of our platform), 64 documentation related PRs had to be submitted and reviewed by the developers. These PRs typically involve manual editing of documentation and hand-crafting of examples by developers, which can be time consuming as well as error-prone. With the deployment of our platform in April 2019, only 20 manual PRs had to be submitted by the developers that year as our platform was able to submit 38 automatic PRs containing machine generated examples, reducing the numbers of PRs developers had to submit by $>$68\% compared to the prior year. 

\section{Lessons Learned and Ongoing Work}
\label{discussion}
Given the benefits and drawbacks of both our neural and feature-based models, we decided to use them both in production. This enabled us to improve both models based on the software owners' feedback. In addition, we learned a few lessons that have guided our ongoing and future work on our example generation platform.

First, we found that the inability of the feature-based model to leverage correlations between the parameters can be problematic in a production system. We faced a few cases, where such examples slipped through the PR review process as they were syntactically correct, but were later caught by end users. This problem did not occur with our neural model, which considers all the parameters when generating values for each parameter. To address this challenge, we are experimenting with ways of combining both models.

Second, we learned that software owners are more tolerant towards examples that have placeholders than examples with incorrect values. Therefore, we are experimenting with a newer version of neural model that can generate placeholders when the confidence is low. For this, we leverage the likelihood that the neural model produces with each prediction. When this likelihood is low, the model falls back to use placeholders or the feature-based model. 

Finally, being a black-box, we also faced challenges tuning our neural model to owners' feedback. 
For instance, when we generated our first automatic PR with the neural model, the software owners asked us to modify the format of all generated names. This meant that we needed to either add a post-processing step or change the formatting of all input parameters and re-train the model. Re-training can be performed quickly, in our case, since our data set is not very large. However, as we try to expand our data set over time, we will look into training separate models, which can modify the style without expensive re-training of the value prediction model.

While in this paper we only discuss the development and deployment of our example generation platform for Azure CLI, the design of our system is generalizable to situations where usage telemetry exists and can be utilized to generate meaningful examples. To demonstrate this, we have also successfully deployed this system to generate examples for Azure PowerShell, another command line tool to manage resources in Azure. If training and usage data is available, our system should also work for generating examples for other command line tools. Similarly our methodology can be used to generate examples for simple API calls targeting cloud services. 
However, our platform in its current form cannot generalize to situations where multiple steps are always required to accomplish a single meaningful task (e.g. scripts). We leave this exploration to future research.

\section{Conclusion}
Up-to-date documentation with many code examples is essential to learning new or updated frameworks and APIs. Yet, official software documentations are often stale and lack sufficient examples. Our work closes this gap by presenting a novel example generation platform that generates up-to-date examples based on usage telemetry. Our evaluation results showed that our examples can help developers by covering all active features with a higher quality than the software owner’s examples. In addition, our example generation pipeline increases software owner’s productivity by $>$68\%. 

An immediate direction for future work is to expand our example generation pipeline to create example scripts (i.e., chaining a series of commands). Another direction is to measure the long-term effect of our platform on the overall quality of Azure CLI documentation. For example, measures can include the amount of time users spend on the online documentation website, the number of documentation related issues reported, or the number of user failures caused by an incorrect combination of command, parameters, or parameter values. Finally, a similar approach can be applied to other tools where usage telemetry is available. We have already deployed the same example generation platform for Azure PowerShell, another command line interface for Azure, to a similar success.



\bibliographystyle{ACM-Reference-Format}
\bibliography{references}


\end{document}